\documentclass[12pt]{article}
\usepackage{graphicx}
\usepackage{amssymb,gensymb}
\usepackage[numbers,sort]{natbib}
\usepackage{hyperref}  % makes toc clickable
\usepackage[outdir=./]{epstopdf}
\usepackage{pifont}
\usepackage{xfrac}

\def\h2o{\ifmmode{{\rm H}_2{\rm O}}\else{H$_2$O}\fi}
\newcommand{\kgmsGa}{\ensuremath{\mathrm{kg\,m^{-2}Ga^{-1}}}}

\begin{document}

\title{Current Theories of Lunar Ice}

\author{
  Norbert Sch\"orghofer \\
  \normalsize Planetary Science Institute, Arizona (norbert@psi.edu), \\
  \normalsize working remotely from Hawaii
}

%\date{\today\vspace{-1em}}
\date{}
\maketitle

\begin{center}
v1 ... February 9, 2025\\
v2 ... \today
\end{center}

\addcontentsline{toc}{section}{Abstract}
\begin{abstract}
  The classical theory of cold-trapped ice on the Moon and some modern theories are reviewed and compared with observational constraints.
  The ``standard model'' for lunar ice posits that ice has accumulated in polar cold traps after the spin axis orientation became small enough for polar craters to be permanently shadowed.
  Its predictions are consistent with major observational constraints.
  Only a few less established observational claims are unaccounted for.
  The text focuses on fundamental theoretical concepts and assumes some pre-existing familiarity with the topic of lunar polar volatiles.
\end{abstract}

\clearpage
\tableofcontents

\clearpage
\section{Ice Storage Processes}

The following ice storage processes have been proposed for the lunar polar regions:
\begin{enumerate}\setlength{\itemsep}{0pt}
  %\item[(i)] implantation of solar wind protons
  \item[(i)] surface cold traps
  \item[(ii)] relic buried ice
  \item[(iii)] subsurface cold-trapping (vapor pumping)
\end{enumerate}

\subsection{Cold-trapping in PSRs}

The Moon has permanently shadowed regions (PSRs) that are so cold that the sublimation rate into the vacuum of space is ``negligible'' \cite{watson61a,watson61b,arnold79}. Here, ``negligible'' means something on the order of 100~kg/m$^2$ ($\sim$10~cm) of ice per billion years, or less than the amount delivered to the PSRs over deep time. The sublimation rate is very sensitive to temperature, so that  a large change in sublimation rate threshold corresponds to only a small change in temperature threshold.
Crystalline (phase Ih) ice at a temperature of 109~K sublimates into vacuum at a rate of 100~\kgmsGa\ (about 10~cm/Ga), and at 114~K the rate is 1000~\kgmsGa\ (about 1~m/Ga), so a 5~K difference corresponds to a tenfold change in sublimation rate. At these temperatures, sublimation is too slow to be measured in the laboratory, but the rates can be reliably extrapolated from measurements at higher temperatures. The main uncertainty is whether the ice is indeed in crystalline form, as at these ambient temperatures it forms in amorphous form \cite{jenniskens98,sack93}. However, the vapor pressure of amorphous ice is only slightly higher than that of crystalline ice. Overall, a threshold of $\sim$110~K for cold-trapping is very defensible.

The idea that craters near the lunar poles may have accumulated ice is more than a century old \cite{goddard1920,urey52}, but it is a pair of papers by Watson, Murray, and Brown in 1961 \cite{watson61a,watson61b} that fleshed out this idea.
They estimated the water sources on the Moon, constructed a model for ballistic hops of water molecules to calculate the fraction of water that ends up trapped, and estimated the area of permanently shadowed regions.
At the time only a few lunar flybys had succeeded and next-to-nothing was known about the polar topography. Their work is now widely considered a seminal contribution, and as will be argued here, it is still an excellent working hypothesis, neither fully proven nor disproven.

\begin{figure}[tb]
  % figures from ~/Papers/Published/MoonSubsDiff21/
  \begin{center}
  \begin{tabular}{c|c}
    \hline
    long-lived & truly stable \\
    \hline
    &\\
  \includegraphics[width=6cm]{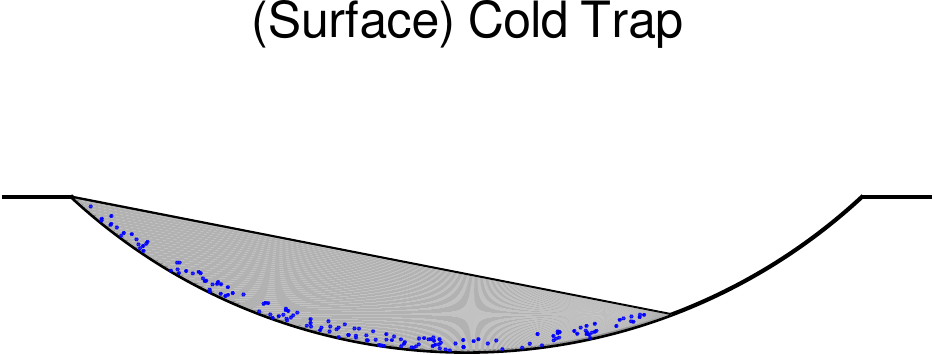}&
  \includegraphics[width=6cm]{storage_fig_coldtrap-crop}\\
  long-lived surface ice & long-term time average of supply \\
                         & exceeds sublimation loss \\
  \hline
  &\\
  \includegraphics[width=6cm]{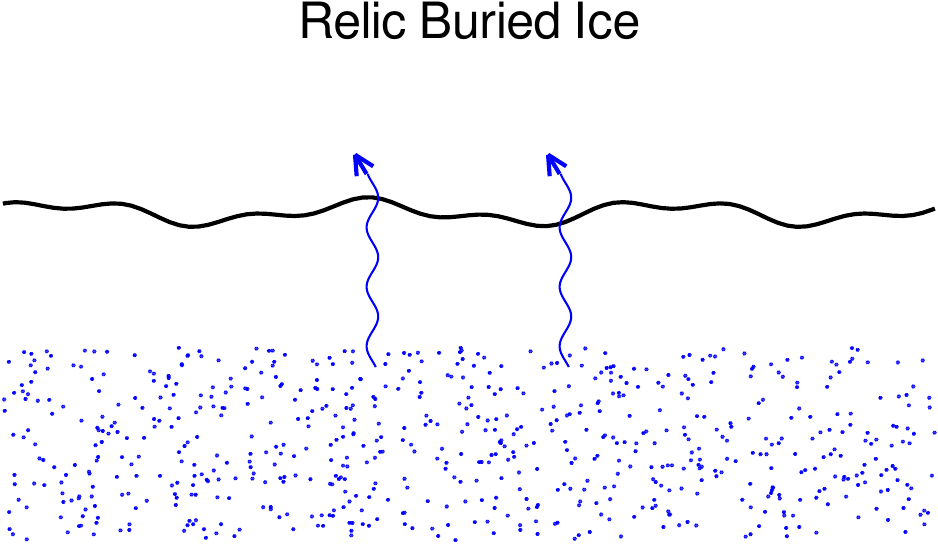}&
  \includegraphics[width=6cm]{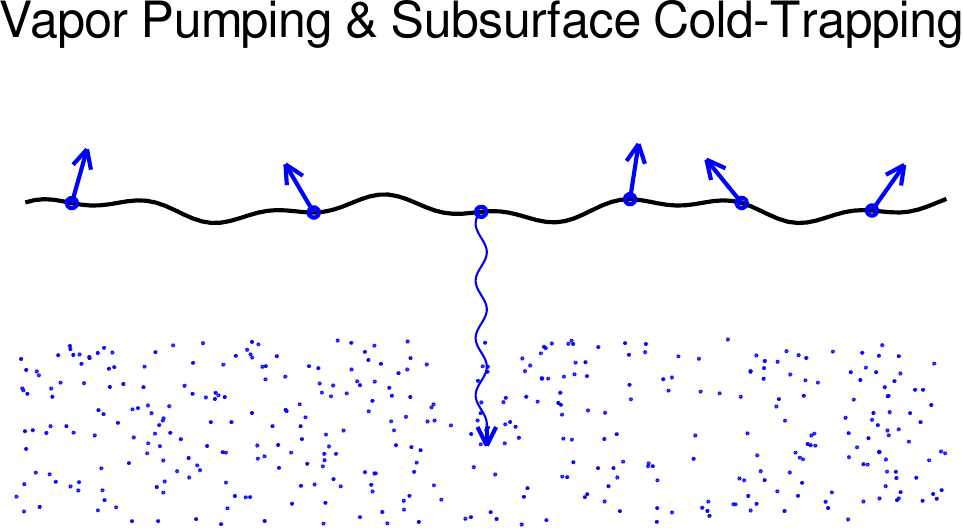} \\
  long-lived subsurface ice & subsurface ice in equilibrium with \\
                            & adsorbed water on surface \\
  \end{tabular}
  \end{center}
\caption{Notions of ice stability on an airless body for three types of ice storage: surface cold traps (gray indicates shadow), relic buried ice, and subsurface cold traps. Adapted from \cite{schorghofer22a}. }
\label{f:notions}
\end{figure}

The original argument given was that ice is long-lived and sublimated only slowly, and the threshold is effectively based on the amount of \h2o delivered to the cold traps. It is often said that ice in PSRs is ``stable'' over geologic time scales, but it is effectively only long-lived rather than in equilibrium with its environment. There is an abstract argument one can give to justify use of the term ``stability'': If a long enough time average is formed (billions of years), the time-integrated supply exceeds the loss and, in this sense, true stability is achieved. This may be hair-splitting about the terminology of ``stable'', but the difference in viewpoint becomes material when considering subsurface ice. Relic buried ice is long-lived but not in equilibrium with the environment, whereas ice cold-trapped in the subsurface is in equilibrium with the surface (Figure~\ref{f:notions}).
%As a digression, appendix B describes how the concept of cold-trapping on airless bodies connects with conventional thermodynamics.

PSRs are defined by shadows, whereas cold traps are defined by temperature (and, more precisely, by time-integrated sublimation loss). Diviner data have made it possible to map cold traps independently of PSRs \cite{paige10ssr,paige10a}. In practice, all lunar cold traps lie within PSRs, and some PSRs, especially those far from the pole, do not act as cold traps because the infrared emission from the sunlit portions heats them to too high a temperature.
The total PSR area is significantly larger than the total cold trap area.
In principle, it is possible to have {\it sunlit} cold traps if illumination is extremely brief and at a grazing angle, but in practice no such case has been identified on the lunar surface. (Even on Ceres, which is much farther from the sun, all cold traps are within PSRs \cite{sgmw24}.)
Some areas receive sunlight for only a brief period during each Draconic year, and these periods are sometimes not covered by Diviner measurements \cite{swm24}.
%Illumination modeling revealed that these {\it apparent} cold traps are not true cold traps.
Table~\ref{t:typesofcoldtraps} lists conceptual types of cold traps.

\begin{table}
  \begin{tabular}{ll}
    \hline
    Surface cold traps \\
    \hline
    Cold trap & standard type\\
    Micro cold trap & spatially unresolved by Diviner, $\lesssim$200~m \cite{hayne21} \\ 
    Apparent cold trap & un(der)sampled by Diviner but briefly illuminated \cite{swm24} \\
    Sunlit cold trap type I & directly illuminated only at low intensity (not observed) \\
    Sunlit cold trap type II & illuminated so briefly thermal inertia prevents heating (not observed) \\
    \hline
    \multicolumn{2}{l}{Subsurface ice preservation} \\
    \hline
    Ice stability region & relic subsurface ice \\
    Subsurface cold trap & cold-trapping in subsurface (vapor pumping)\\
    \hline
  \end{tabular}
  \caption{Types of cold traps and pseudo cold traps.}
  \label{t:typesofcoldtraps}
\end{table}

At very small scales, maybe centimeters, lateral heat conduction suffices to warm mini-PSRs beyond the cold trapping threshold \cite{bandfield15,hayne21}. Lunar fines have extremely low thermal conductivity, about $10^{-3}$~Wm$^{-1}$K$^{-1}$ and less \cite{cremers75}. (For comparison a solid piece of dry rock has a conductivity of about 4~Wm$^{-1}$K$^{-1}$, a difference of 3\sfrac12 orders of magnitude.) The low heat transfer is due to the vacuum, only tiny contact points between grains, and diminishing radiative heat transfer at low temperature. Cold fine-grained material in vacuum is the best natural thermal insulator. Temperature gradients can be huge, vertically and horizontally, and PSRs even as small as a centimeter or even less may be able to remain cold enough \cite{bandfield15,hayne21}.
However, such small craters are statistically very young and therefore could not have accumulated much ice even under favorable circumstances \cite{costello24,rubanenko24}.
%(Another theoretical aspect of micro cold traps is described in Appendix~C.)

Cold traps are defined by negligible (or sufficiently slow) sublimation loss. They are still subject to space weathering, such as all-sky Lyman-$\alpha$ radiation, sputtering from charged solar wind plasma, and micrometeoroid bombardment which erode ice \cite{farrell19}. Pessimistic weathering loss rates are on the order of 0.2~nm/yr \cite{lanzerotti81,morgan91,farrell19}. Water delivery needs to be sufficiently fast to overcome these loss processes, otherwise no ice will accumulate.
From that perspective, episodic delivery is more effective than continuous delivery; subsequent impact gardening can mix the regolith and protect the ice from unending space weathering.

\subsection{Cold Silicate Hypothesis}

The ``Cold Silicate Hypothesis'' was formulated primarily by Larissa Starukhina at Kharkov University around the year 2000 \cite{starukhina00,starukhina01}, after the detection of hydrogen excess at the lunar poles in 1998 \cite{feldman98} and the discovery of radar-bright regions on Mercury in 1992 \cite{slade92,harmon92}.
She argued that implanted solar wind hydrogen forms OH and accumulates in very cold regions on the Moon, whereas in warmer regions OH is lost after diffusion, and that the radar properties of silicates at low temperature are uncertain. Hence, neither of the two discoveries necessarily implies the presence of water ice.

With the LCROSS mission, which detected molecular water in a lunar PSR \cite{colaprete10}, and the MESSENGER mission to Mercury \cite{lawrence13,neumann13}, the hypothesis has been shown to be wrong. However, the idea of temperature-controlled OH accumulation may be applicable in a different way, in form of a latitude-dependent hydroxyl veneer that has since been observed \cite[e.g.,][]{pieters09,clark24}.
It may also apply to the rate of water diffusion inside lunar grains after water is generated by solar wind on the amorphous grain rim \cite{he23}, thus potentially causing a latitude dependence for grain-internal water. Diffusion rates are governed by Boltzmann factors, $\exp(-E_{\rm act}/k_B T)$, and their strong temperature dependence can result in geographically narrow transitions.

\subsection{Vapor pumping}
\label{sec:pumping}

The theory of subsurface cold-trapping involves an interesting concept: vapor pumping due to temperature cycles which accelerate the transport from the surface into the subsurface.
The concept of ``vapor pumping'' originated from Mars research \cite{mellon93}, but the term was introduced only later. On Mars, water vapor from the atmosphere can be sequestered as ice at depth. This occurs when the time average of the atmospheric vapor pressure exceeds the vapor pressure at the subsurface ice table. Subsurface ice can be in vapor-exchange equilibrium with the atmosphere, without presence of ice on the surface.
It is a finite-amplitude effect, meaning it does not occur without a temperature amplitude.
A remarkable aspect of this phenomenon is a stark contrast between surface and subsurface ice concentrations (as observed on Mars). In fact, ice can be sequestered even if there {\it never} is any frost on the surface and the atmosphere is {\it never} saturated \cite{schorghofer07b}.

The vapor pumping phenomenon can also occur on an airless body, if there are mobile adsorbed water molecules on the surface which serve as source for the sequestration \cite{siegler11,sa14,reiss21,schorghofer22a}. Initially, this was demonstrated for the Moon with an episodic ice cover as source \cite{schorghofer07a}. However, once the ice cover is removed, the sequestered ice quickly disappears again. For that reason, short-lived thermal events are unlikely to result in long-term subsurface ice storage \cite{siegler11,siegler16}. With a continuous source, the sequestration proceeds indefinitely, and ice will steadily accumulate in the subsurface. Using known surface temperatures and assumed water supply rates, ice sequestration can occur over polar areas on the Moon far larger than the area covered by PSRs \cite{sa14,sw20,swm24}. A hallmark of this phenomenon is that ice accumulation occurs at a specific depth, where ice becomes ``stable'' (sublimation loss is lower than the supply), whereas no accumulation occurs at shallower depths. Vapor is cold-trapped in the {\it sub}surface, and vapor pumping is the process which supplies the subsurface cold traps. The depth is essentially set by the thermal skin depth. As a rule of thumb, subsurface cold-trapping occurs where subsurface temperatures are continuously below a threshold of $\sim$110~K. This is effectively the same criterion as for surface cold traps, with the surface peak temperature replaced by subsurface peak temperature.

Subsurface cold-trapping requires a quasi-continuous source of water molecules on the surface, which could be delivered by an exosphere or perhaps generated in-place by solar wind. The process has very low yield; only a small fraction of the water adsorbed on the surface is sequestered \cite{schorghofer22a}. However, if vapor pumping does occur, it will occur over large swaths outside of cold traps. It also occurs inside of cold traps, but at an uninterestingly slow rate.

The term ``ice stability region'' (ISR) is often used to refer to buried relic ice, i.e., ice protected from sublimation by an overlying layer. This is distinct from sub-surface cold-trapping, which is even more deserving of the name ``stability'' (Fig.~\ref{f:notions}). Both are regions where ice will survive sublimation loss, but they are due to different processes and expected in different (but overlapping) geographic areas.

A weak form of vapor pumping is the ``adsorbate pump'', which results in enhanced concentrations of adsorbed \h2o instead of ice \cite{sa14,schorghofer25}, and may act globally.

\section{Sources and Transport over Time}

\subsection{Polar wander}

The existence of PSRs is a necessary condition for polar ice, so polar wander (true or apparent) is of great consequence. Old cold traps should have captured far more ice than young cold traps.

\begin{figure}[tb]
  % figures from ~/Moon/PSRs/Figs/
  \centering
  \includegraphics[width=8cm]{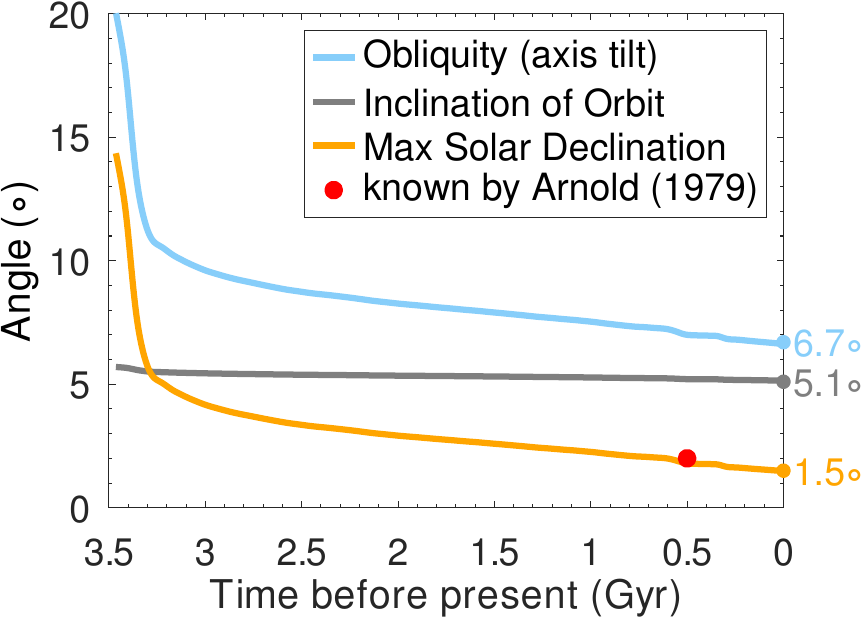}
  \caption{Evolution of the lunar spin axis after the Cassini State Transition \cite{schrufu23}. Water ice only accumulated after the first polar cold traps formed. }
  \label{f:obliq_vs_time}
\end{figure}

William Ward \cite{ward75} realized that the tidally locked state the Moon currently occupies did not exist when the Moon was much closer to Earth, and the Moon, which formed close to Earth, must have undergone a spin axis reorientation as it transitioned from one Cassini State to the other.
James Arnold \cite{arnold79} (the same James R.\ Arnold who worked with Libby on the development of radiocarbon dating) realized the significance the evolution of the lunar spin axis has for lunar polar ice. At the time, only one data point from 0.5~Ga ago was available for the length of Earth's day, which constrains the tidal evolution of the Earth-Moon system. Based on that one uncertain data point, Arnold estimated that a typical age for the PSRs is only 2~Ga.
More modern analysis, with additional and more certain data \cite{farhat22}, upheld this analysis \cite{schrufu23} (Figure~\ref{f:obliq_vs_time}).
Although the Cassini State transition occurred a long time ago (maybe 4.1 Ga; the exact time is still uncertain), the relaxation of the spin axis toward its current orientation was slow. The maximum solar declination had twice the current value (3\degree\ instead of 1.5\degree) only 2.1 Ga ago \cite{schrufu23}, making the PSRs much smaller. Almost no PSRs existed more than 3.4~Ga ago, during the Late Imbrian. In other words, accumulation of ice occurred only over the last few billion years. Early in the history of the Moon, outgassing and large impacts were far more common \cite{cannon20b}, but any ice cold-trapped from that period was lost during the Cassini State transition.
%At one point in time, the lunar spin axis was close to the orbital plane.

The large amplitude of true polar wander due to the Moon's outward migration makes it unlikely that polar wander due to internal mass redistribution, which has also been suggested \cite{siegler16}, played a significant role over the last few billion years.

\subsection{Sources of water}

Briefly, the following potential sources of water have been considered.

{\it Impacts of comets and hydrated asteroids (meteoric \h2o).}
Comets are ice-rich and many asteroids also contain significant amounts of water. The retention rate depends greatly on the impact velocity, so hypervelocity impacts, common for cometary orbits, deliver less water than slower impacts, common for asteroidal orbits.
Impacts are an episodic source, which can create a temporary dense atmosphere.
The impactor mass flux is dominated by the largest object, which is subject to small number statistics.
One can argue that Mercury got lucky and experienced a major water delivering impact and that is why Mercury's cold traps have so much more ice.
%Impacts are thought to the largest source of lunar polar water \cite{cannon20b}.

{\it Outgassing.}
Volcanic activity peaked 4--2~Ga ago, but may have episodically continued afterwards \cite{wilcoski22}.
Ancient lunar volcanism may have released enough water to produce a fully collisional atmosphere, but most outgassing events probably resulted only in collisionless exospheres \cite{needham17,head20}.

{\it Solar wind generation.} Significant progress has been made on this topic over the past decade, both from laboratory experiments and analysis of new return samples.
The flux of solar wind proton at 1~AU is about $2\times 10^{12}$~m$^{-2}$s$^{-1}$ \cite{sznajder23}, and due to the lack of a global magnetic field most of them reach the lunar surface. Per lunation, this amounts to $2 \times 10^{12} \times 86400\times 29.53/4 = 1.3$ protons per nm$^2$ on a global average. For comparison, a monolayer of \h2o ice has 10 \h2o molecules per nm$^2$.

Solar wind can generate {\it molecular} water on the lunar surface. A long-proposed process was reduction of ferrous iron by solar wind protons \cite{housley74}. Now a different process has been established, namely recombination of solar-wind generated OH \cite{jones18,jones20a,zhou22}. The reaction MOH+MOH$\to$MOM+\h2o(g) produces molecular water \cite{jones18,jones24}, where ``M'' stands for a metal atom. The activation temperature for this process is nearly 400~K, a temperature not reached on the lunar surface. Micrometeoroids could provide temperature spikes that activate the OH-recombination on the lunar surface \cite{zhu19}, but on a global average this would still yield a low production rate. Overall, solar wind generation processes, although present, are expected to be a negligible contribution to polar ice on the Moon.
The activation temperature is surpassed on Mercury, where solar wind generation of \h2o is estimated to have been able to produce 10\% of the observed polar ice deposits \cite{jones20mercury}.

%Solar wind generated water is created continuously and can steadily supply a water exosphere.

Analysis of return samples from the Chang'e~5 mission has established that OH in lunar rocks is largely generated by solar wind \cite{liu22,zhou22,he23}.
However, this does not imply that ice in the PSRs is similarly derived from solar wind, as this internal hydroxyl is not released.

{\it Dielectric breakdown caused by solar energetic particles (SEPs).}
Huang et al.\ (2021) proposed that molecular water can be generated by dielectric breakdown directly in PSRs \cite{huang21a}. Energetic solar events create strong electric fields on the lunar surface and the subsequent dielectric breakdown can break the chemical bonds of regolith grains and expose the oxygen atoms to react with the hydrogen implanted by solar wind.

Impacts and outgassing are thought to be the largest sources of the ice that ends up in the lunar polar regions and both of these sources peaked early. For that reason the age of cold traps (Figure~\ref{f:obliq_vs_time}) is critical for the amount of polar ice that has accumulated.
The estimate by Arnold \cite{arnold79}, which has still not been updated, is 1-10\% ice in the top 2 meters, where the depth arises from estimated impact mixing depth over two billion years.

\subsection{Exospheric Transport Hypothesis}

The gravity of the Moon and Mercury is high enough to retain molecules gravitationally, assuming the launch (desorption) velocities are at thermal speed. This gives rise to gravitationally bound and surface-bounded exospheres that consist of ballistic trajectories.

One component in the original work by Watson et al. \cite{watson61b} is that water can be transported from any location on the lunar surface to polar cold traps through ballistic lateral flow. Molecules migrate until they are destroyed by photo-dissociation or land in a cold trap.
This process can concentrate water in the polar regions, in the sense that the flux delivered to cold traps (per area) is much higher than the average flux emitted from the lunar surface (per area). For the physical parameters of the Moon, the process is especially effective \cite{schorghofer17c}. 

Exospheric transport has been established for other species, like argon, but \h2o is more reactive and it has been argued that \h2o will not survive contact with the pristine space-weathered lunar surface \cite{hodges91,hodges02}.
According to this view, most water molecules that land are assimilated into the regolith \cite{hodges91} and ultimately removed by sputtering or meteoroid impacts at superthermal speeds \cite{hodges02}.
Lengthy discussions can be had about this, but the bottom line is that the hypothesis of a water exosphere supplying the PSRs \cite{watson61a,watson61b} has so far neither been proven nor disproven.

Theory has something to say about whether all polar cold traps should fill at the same rate. A global water atmosphere would supply all of them equally. For an exosphere this was less clear, but the question has been settled by now. 
Unless the water source was very close to an individual PSR, delivery to the polar region is predicted to be fairly uniform \cite{schorghofer14c,prem18}, so the geographic location of a cold trap barely matters. The average ballistic hop length is hundreds of km, larger than any PSR, so there is no rain shadow effect. (But seasonally shadowed regions can become that large \cite{kloos19}.)
%An early exosphere model result that suggested otherwise \cite{moores16} has been corrected by the same authors \cite{kloos19}.
%In conclusion, unless the water source is very close, cold-trapping from an atmosphere or an exosphere are both predicted to be fairly uniform, so the geographic location of a cold trap barely matters.

At temperatures well below the cold trapping threshold of $\sim$110~K, sublimation loss is negligible even relative to the amount delivered. Hence, the amount of ice that has accumulated should be unrelated to temperature, other than an indirect correlation of temperature with the duration of permanent shadow (the age of PSRs). There is no reason to expect cold cold traps harbor more water ice than warm cold traps.

Surface cold traps could have captured \h2o from a collisional atmosphere or a collisionless exosphere.
Ice could have accumulated in PSRs by either of these two transfer paths. For subsurface ice, on the other hand, the distinction between continuous and episodic supply is of greater consequence.
Relic buried ice would start with an episodic delivery event and it cannot be recharged after burial. The opposite is the case for ice formed by subsurface cold-trapping; it requires quasi-continuous delivery and cannot be created by a few episodic events.

\section{Observational Constraints}

\subsection{Ice in PSRs}

The discovery of radar-bright deposits in the polar regions of Mercury in 1992 \cite{slade92,harmon92,butler93} (re)motivated the search for ice on the Moon.
A radar investigation by the Clementine spacecraft initially reported evidence for polar ice \cite{nozette96}, but subsequent analysis of the same data and additional radar measurements arrived at the opposite conclusion \cite{stacy97,simpson99,campbell03,eke14}.

Lunar Prospector (1998--1999) carried a nuclear spectrometer, LPNS. Cosmic rays generate $\gamma$-rays and neutrons that are passively measured by the detectors in orbit around the Moon. Fast neutrons are quickly moderated when they collide with nuclei of equal mass (i.e., hydrogen), and the method is especially sensitive to the H-concentration. Passive neutron spectroscopy senses about the top meter of the surface.
LPNS discovered: (i) a hydrogen excess at both polar regions of the Moon relative to the rest of the lunar surface, and (ii) in some areas the hydrogen-rich layer lies beneath a layer with no excess hydrogen \cite{feldman98,feldman00a,feldman01,lawrence06,miller14}. The horizontal resolution of this technique was low, but if the hydrogen excess is assumed to be only in the PSRs, the abundance is about 1.5~wt\% \h2o-equivalent.

Neutron spectroscopy was repeated by a competing group from Russia with the LEND instrument on LRO. They arrived at the same conclusions, namely a polar hydrogen excess and burial \cite{mitrofanov10}. Hence, these facts are well-established. The LEND instrument was designed to achieve higher spatial resolution using collimators, and this has been the subject of controversy \cite{lawrence10,lawrence11comm,eke12,miller12,teodoro14,eke14}, which distracts from the fact that LEND confirmed the LPNS results \cite{mitrofanov12,boynton12}.
% true for both omnidirectional LEND sensor (mitrofanov12) and smoothed version of collimated LEND data (boynton12)

Other than implantation of solar wind protons, the only counterargument given against the interpretation of polar hydrogen excess as water is a claim that a local deficit of calcium would have the same effect on neutron counts \cite{hodges02b}, but that idea found no traction among nuclear physicists \cite{lawrence06}.

The LCROSS impact proved that molecular water is indeed present in a PSR \cite{colaprete10}. A fraction of $\sim$6\% was observed in the plume, representing the average over the volume excavated by the impact.
This abundance is consistent with the classical prediction by Arnold \cite{arnold79}.

ShadowCam, designed to see the surface of PSRs illuminated only by sunlight scattered from the crater walls, did not find bright albedo patterns \cite{robinson24lpsc}. The threshold for reflectance contrasts that can be ascertained may be as high as 30\% (above background reflectance) due to only indirect illumination \cite{martin24}.
The surfaces of PSRs are similar to normally illuminated slopes and plains \cite{basilevsky24b}.
A few features were discovered that are consistent with {\it sub}surface ice. Lobate-rimmed craters are a morphology indicative of water ice in the target material \cite{robinson24lpsc,basilevsky24b,basilevsky25}.
And some craters within PSRs have unusual pockmark morphology that may have resulted from devolatization \cite{danque24abs}.
%The ShadowCam observations are consistent with the results from nuclear spectroscopy, because the latter have long suggested burial.

LPNS, LEND, LCROSS, and ShadowCam were specifically designed to look for water ice. Several other instruments were used to pry information about lunar volatiles. Table~\ref{t:listofobservations} provides an overview.
Measurements of the reflectivity of the optical surface at various wavelengths (LOLA at 1.064$\mu$m, LAMP in the UV, and M$^3$ from 0.43 to 3.0~$\mu$m) suggest a minor fraction of the surface area of PSRs is covered by water. This strengthens the case for ice in the lunar cold traps. Unfortunately, the three maps show no significant correlations with one another, so at least two of them do not represent exposed \h2o. The geographic distributions are also unrelated to any other known physical properties (except a correlation with surface temperature).

At the high resolution of ShadowCam (1.5~m/pixel), no bright ejecta (or only very few apparently bright features) are seen.
The LCROSS impact site has been imaged 15 years later by ShadowCam, and no part of it appears bright \cite{fassett24lcross}.
The bulk concentration of ice at depth may only be a few percent, so ejecta contain at most that percentage of ice, causing reflectance contrasts below ShadowCam's sensitivity limit.

\begin{table}
\begin{tabular}{llc}
  \hline
  Instrument/Observation & Reproduced or Confirmed & Confidence \\
  \hline
  {\it Methods that sense to depth} \\
  \hline
  radar, Clementine bistatic \cite{nozette96} & contradicted by \cite{simpson99} &  \ding{55} \\
  radar, Arecibo, no massive ice \cite{stacy97} & lack of massive ice confirmed \cite{campbell03,neish11} & \checkmark \\
  LPNS, H abundance, 1.5\% WEH \cite{feldman00a,feldman01} & confirmed by LEND & \checkmark \\
  LPNS, H burial \cite{feldman98,feldman00a,feldman01} & confirmed by LEND & \checkmark \\
  LEND high resolution \cite{mitrofanov10,sanin17} & disputed \cite{lawrence10,lawrence11comm,teodoro14}  & ? \\
  LCROSS, \h2o \cite{colaprete10,gladstone10} & no, but undisputed & \checkmark \\
  Mini-RF, boulders or ice \cite{spudis10,spudis13} & boulders \cite{fa24abs,verma25} & resolved \\
  crater depth:diameter statistics \cite{rubanenko18} & \cite{sathyan24abs} with mixed results & ? \\
  \hline
  {\it Methods that sense the surface} \\
  \hline
  SELENE-TC, no frost in Shackleton \cite{haruyama08} & \cite{zuber12,haruyama13,mahanti25} & \checkmark \\
  LAMP, UV, 0.1--2\% frost or OH \cite{gladstone12} & partially \cite{hayne15b} & ? \\
  LOLA, 1064nm albedo, frost \cite{fisher17} & may be space-weathering instead & \ding{55} \\
  M$^3$ in PSRs 1.3, 1.5, 2.0 $\mu$m frost  \cite{shuaili18} & -- & ? \\
  ShadowCam, $\lesssim$30\% frost \cite{mahanti23,robinson24lpsc} & no, but undisputed & \checkmark \\
  \hline
\end{tabular}
\caption{Observations of polar ice}
\label{t:listofobservations}
\end{table}

According to the observational results summarized in Table~\ref{t:listofobservations}, a coherent picture of lunar ice emerges:
Methods that sense depth indicate a moderately low percentage of ice, but not large concentrations, within the depth they are sensitive to.
Methods that sense only the optical surface have produced some evidence for regional ice exposures, but little coherent evidence, and ShadowCam provided a compelling upper bound on the presence of frost on the surface of PSRs.

{\it Is it possible that there is no ice on the Moon at all?}
``Ice'' refers to \h2o in bulk phase, not just adsorbed \h2o or internal water. A monolayer of \h2o corresponds to about 150~ppm in bulk concentration, using the average specific area of lunar fines, so a few monolayers can add up to close to 0.1\%. Internal concentrations in return samples are typically a few tens to hundreds of ppm \cite{mccubbin15rev}, consistent with global measurements from nuclear spectroscopy, and 6-$\mu$m-spectroscopy suggests local concentrations reach several hundreds of ppm and more \cite{honniball21} (but also see \cite{wilk25}), so the internal water content amounts to possibly another 0.1\%. 
The LPNS abundance is 1.5$\pm$0.8~wt\% \cite{feldman00a}. With a higher hydrogen content for the surrounding area, the excess is lower, $\sim$0.5\% \cite{elphic07}.
The LCROSS detection is 5.6$\pm$2.9\% \cite{colaprete10} relative to an excavation volume that had to be estimated and could have been contaminated if fuel was left in the impacting rocket stage. And it was only a single impact experiment. At the lower end of the stated uncertainty, there would be only 2.7\% at the LCROSS site.
Abundances of 1\% and above suggest the presence of ice, but the observational evidence for ice is admittedly not overwhelmingly strong.

More generally, surface phenomena like adsorption or surface hydroxylation do not provide enough H, OH, or H$_2$O abundance, despite the extremely high specific surface area of lunar fines. One monolayer of \h2o over a specific surface area of 500~m$^2$/kg is only 0.15~g/kg (0.015 wt\%). High H-content points toward a bulk phenomenon, such as ice, internal water, or, maybe, ``deep'' hydroxylation in 100~nm thick grain rims.

\subsection{Ice outside of PSRs}

LEND has identified H-excess outside of PSRs. These are called ``neutron suppression regions'' (NSRs) \cite{mitrofanov12,boynton12,sanin17}, due to the reduced number of epithermal neutrons, which correlate negatively and monotonically with the H-concentration. Solar-wind implantation and micro cold traps do not provide enough bulk abundance to explain these measurements.
Two processes have been proposed to explain how ice may be able to survive sublimation loss in these regions: burial of ice, which survives as relic ice beneath that layer, and subsurface cold-trapping, enabled by vapor pumping through temperature cycles.
Both are illustrated in Figure~\ref{f:notions},

Relic buried ice is protected by an overlying layer and could last for billions of years for temperatures up to about $\sim$140~K \cite{schorghofer08a}. But how does the ice survive before it becomes buried? Icy ejecta could originate from a nearby cold trap. Or ice could have formed over a large area after a major delivery event and then be quickly buried by dry ejecta. Both of these scenarios involve local events and many exploratory drill holes may come up dry. Self-burial by a sublimation lag represents another hypothetical scenario, but no evidence of lags has been observed on the Moon. (On Mercury, dark lags are thought to exist \cite{chabot18rev}, presumably formed by radiation-processing of carbon-rich materials.)

Vapor pumping and subsurface cold-trapping could occur over large areas, as long as there is a supply of \h2o over extended periods. A monolayer of \h2o per year would easily suffice, which corresponds to about 30~cm/Ga. The predicted depth to ice is governed by the thermal skin depth, $\sim$5~cm \cite{schorghofer22a}, which is comparable to the burial depth that has been inferred from neutron spectroscopy ($\sim$3--30~g/cm$^2$ \cite{lawrence11}). The few comparisons that have been made between the locations of NSRs and vapor-pumping conditions are favorable \cite{glaser21}. Many quantitative aspects remain to be worked out, but vapor-pumping would provide a natural explanation of subsurface ice outside of PSRs, both in terms of abundance and burial depth. The presence of ice sequestered in subsurface cold traps outside of PSRs could be verified with a single borehole at an appropriate location. 

\begin{table}
  \begin{tabular}{lcc}
    \hline
    Theory                  & Abundance & Burial \\
    \hline
    Solar wind implantation & insufficient & no \\
    Micro cold traps  & insufficient & no \\
    Relic buried ice (``ice stability regions'')  &  ?  &  yes \\
    Vapor pumping \& subsurface cold-trapping & ? & yes $\sim$10~cm \\
    \hline
  \end{tabular}
  \caption{Proposed explanations for H-excess outside of large cold traps.}
  \label{t:outside}
\end{table}

\subsection{Water exosphere}

\begin{table}
\begin{tabular}{lll}
  \hline
  Instrument & Observation & Notes \\
  \hline
  Apollo LACE \& CCGE, surface \cite{hoffman73cospar} & $<10^{5}$ cm$^{-3}$ & nighttime total \\
  ARTEMIS \cite{halekas13} &  $<1.7\times 10^{4}$ cm$^{-3}$ OH \\
  Chang'e 3 (UV telescope) \cite{wang15} & $<1\times 10^{4}$ cm$^{-3}$ OH \\ 
  CHACE-1, descent  \cite{sridharan15corr} & $\sim 10^{6}$ cm$^{-3}$ \\
  LADEE, equatorial orbit \cite{benna19} & 23 cm$^{-3}$ \h2o or OH & challenged by \cite{hodges22} \\
  SELENE spectroscopy \cite{ohtake24} & vapor plumes \\
  CHACE-2, polar orbit \cite{chakraborty25} & $\sim$5000 cm$^{-3}$ \h2o \\
  \hline
\end{tabular}
\caption{Observations of lunar water exosphere. LACE, CHACE, and LADEE are mass spectrometers.}
\label{t:exoobservations}
\end{table}

Several instruments have detected water in lunar orbit (Table~\ref{t:exoobservations}). However, the claimed abundances contradict one another. Moreover, mass spectrometers suffer from an OH/H$_2$O ambiguity, because reactive OH can convert to \h2o on the interior walls of the instrument. CHACE-2 should be less affected by this ambiguity, and Ref.\ \cite{chakraborty25} reported about equal densities for \h2o and OH. Overall, the density of the lunar water exosphere is uncertain, because of contradictory observational claims, so the Exospheric Transport Hypothesis for \h2o can still be challenged. Observations of a global \h2o exosphere would prove that water can be continuously transported to the lunar poles.

\subsection{Supervolatiles}

The theory of cold-trapping also applies to chemical species more volatile than \h2o, only over smaller areas \cite{berezhnoy03,landis22,sw24}.
Among the species abundant in comets (e.g., H$_2$O, CO$_2$, CO, CH$_3$OH \cite{bockelee17}) or outgassed from planetary interiors (e.g., H$_2$O, CO$_2$, SO$_2$), water has the lowest vapor pressure, which explains its relative abundance in the solar system \cite{watson61b}.

The location of the LCROSS impact site was exceptionally cold and indeed harbored supervolatiles. Eight additional chemical species were detected, including hydrocarbons and sulfur-bearing species \cite{colaprete10}.
The observed LCROSS abundances, relative to \h2o, do not follow the expected abundances in potential sources, the rank in volatility, or a combination thereof \cite{berezhnoy12}. However, the absolute abundances of the supervolatiles detected in the LCROSS plume are low, and no errorbars for them have yet been published. The unexplained ranking of supervolatile abundances could change within these uncertainties.

\section{Standard Model of Lunar Ice}

\subsection{Classic and modern view of lunar ice}

The classic theory of lunar ice with moderate modifications is here referred to as the ``standard model'' of lunar ice. Table~\ref{t:theories} lists key elements of the classic and modern view of ice on the Moon. ``Classic'' refers to Watson et al.\ (1961) \cite{watson61a,watson61b}, Arnold (1979) \cite{arnold79}, and earlier works. The modern view came about starting with the first nuclear spectroscopy measurements by Feldman et al.\ (1998) \cite{feldman98}.

\begin{table}[tbh]
  \begin{tabular}{lll}
    \hline
    & Classic   & Modern \\
    \hline
    Location:   & PSRs \cite{goddard1920,urey52,watson61a,watson61b}  &  cold traps \cite{paige10a} \\
    Transport:  & exospheric or atmospheric \cite{watson61a,watson61b,arnold79} & unchanged \\
    Sources:    & outgassing, comets, asteroids, solar wind   & outgassing, asteroids \\
    PSR history: & Cassini State transition; no ancient ice \cite{arnold79} & upheld \cite{farhat22,schrufu23} \\
    Concentration: & 5--50 g/cm$^2$, 1--10\% \cite{arnold79} & upheld, one to tens of wt\% \\
    Stratigraphy: & --                   & ice is buried \cite{feldman98} \\
    Ice destruction: & photodissociation, solar wind, impacts \cite{arnold79}   & + hidden by mass diffusion? \\
    Other:      & ice only in PSRs       & excess H outside of PSRs \cite{mitrofanov10} \\
    \hline
  \end{tabular}
  \caption{Comparison of the classical theory of lunar ice with the modern perspective.}
  \label{t:theories}
\end{table}

We have moved from PSRs to cold traps, a distinction which was not made in the classical works but is a straightforward refinement. PSRs have been mapped using raytracing of laser altimeter based topography \cite{mazarico11,mazarico18}. Cold traps have been mapped with Diviner, a radiometer on LRO \cite{paige10ssr,paige10a}.

Comets are now thought to be a negligible source of polar ice \cite{cannon20b}; the mass flux of hydrated asteroids and meteoroids is higher, and the fraction retained is also higher compared to hypervelocity comets.

As described above, knowledge of solar wind processes has increased a lot. But ultimately, solar wind generation processes, although present and dominant on the sunlit lunar surface, are expected to be a negligible contribution to polar ice on the Moon.

The role of the spin axis evolution due to Moon's outward migration is far better supported by data than it was in 1979. The ice concentrations estimated by Arnold are 1--10\%, perfectly consistent with the measurements by LPNS ($\sim$1.5\%) and LCROSS ($\sim$6\%). If this continues to hold true, the classic theory of lunar ice as formulated by Arnold \cite{arnold79} is headed toward a stunning victory.
Neither the classical nor modern theory predict ``massive'' ice deposits (tens of percent).
Only ignoring the evolution of the lunar spin axis would predict such high abundances.

According to the standard model, the history of lunar ice is as follows:
Cold traps start to form roughly one billion years after the Cassini State Transition. They captured some of the water emitted from outgassing and from hydrated impactors, processes that were slowing down but still going on. These episodic events fed the cold traps, whereas continuous delivery was at most a minor part. The ice was subsequently mixed by impacts, and surface ice exposures in PSRs are erased by space weathering. Today we see a percentage of ice within the top few meters, and little ice on the surface. Burial occurred due to impact mixing and landslides.

Destruction of surface ice in PSRs by solar wind, photodissociation (from starlight or interstellar Lyman-$\alpha$) and impacts have all already been considered in the 20th century, and the estimates have not changed dramatically, e.g., 0.1~nm/yr estimated in 1991 \cite{morgan91} versus 0.3~nm/yr estimated in 2019 \cite{farrell19}. That said, the 1991 destruction rate by all-sky Lyman-$\alpha$ has been revised downward \cite{gladstone12}.
On dwarf planet Ceres, ice has survived near its surface even after billions of years of impact gardening and impact heating \cite{prettyman21}, so impacts apparently do not devolatilize airless bodies in a major way. Landslides may have buried ice on crater floors, which is not a destructive process, but it hides the ice from remote nuclear spectroscopy and makes it more difficult to access.

So far, the classic theory, plus burial, is consistent with observations and can be referred to as the ``standard model'' of lunar ice. The same model, only with a different pole orientation history and source rates, also does well in explaining the polar ice deposits on Mercury. That said, the standard model is far from proven.  There is a need for {\it definitive} measurements of lunar polar volatiles. In particular, the most likely locations to have high concentrations of ice, old cold traps, have never been visited by a lander.

\subsection{Challenges to the Standard Model}

%Ice outside of PSRs goes fundamentally beyond the classic ideas, and must be due to other processes, such as those listed in Table~\ref{t:outside}
Not predicted by the standard model is the hydrogen excess observed outside of cold traps, although, like most observational claims about solar volatiles, this is not yet fully established. This requires other processes, such as an extraordinary accumulation of solar wind protons or vapor pumping that sequestered \h2o (Table~\ref{t:outside}). These polar sunlit areas are more easily explored than large PSRs, so we may get additional observational constraints from in-situ experiments sooner for them than for the deep dark craters.

Due to the evolution of the lunar spin axis, old cold traps should have more ice than young cold traps. The LCROSS impact site is in a young PSR (it became permanently shadowed only $\sim$0.9~Ga ago) \cite{schrufu23,fassett24lcross}, but has a high excess of hydrogen based on LEND data \cite{sanin17,litvak24}, which was one of the reasons the location was chosen for the LCROSS experiment. Moreover, the LEND data do not suggest the oldest PSRs (e.g., Shackleton Crater) have the most ice.
These observations fundamentally contradict the idea that fewer cold traps existed in deep time due to the history of the lunar spin axis orientation. However, the high-resolution LEND data have not yet been independently confirmed, so there is room to argue that the locations with the highest ice concentrations are in fact elsewhere. Alternatively, the top half meter may have been subjected to regionally varying modifications, which masks the ice content at greater depths.
This may be called the ``heterogeneity problem'' of the standard model.

Impacts and mass wasting, processes that remove ice from the surface, are heterogeneous in nature, so near-surface ice concentrations can be expected to be heterogeneous. Shackleton Crater, one of the oldest PSRs, may have experienced a massive regolith slide from the inner walls of the crater, which would have covered up the ice beyond the sensing depth of nuclear spectroscopy \cite{basilevsky12rev}.
Shackleton has an unusually high depth-to-diameter ratio, so its interior experiences more self-heating than most lunar craters. Although Shackleton may contain the oldest PSR, it might be far from the oldest cold trap.

Geographic patchiness of exposed ice was inferred from surface reflectivity measurements by LOLA, LAMP, and M$^3$ at various wavelengths within PSRs (Table~\ref{t:listofobservations}). These measurements resulted in three mutually contradictory maps, if interpreted as \h2o surface exposures. None of the three maps agrees with the LEND-inferred geographic distribution either. That said, small-scale heterogeneity ought to be expected due to sporadic destruction mechanisms, but regional heterogeneity (some old cold traps appear to have less ice than some young cold traps) is more challenging to explain.

\subsection{Predictions and key measurements needed}

The standard model of lunar ice, if correct, results in a number of predictions:
\begin{itemize}\setlength{\itemsep}{0pt}

\item All cold traps older than the cold trap at the LCROSS impact site (PSR age $\sim$0.9~Ga) should have accumulated ice, and younger ones may have as well. 
  
\item The abundance of ice (column-integrated) should vary with the age of the cold trap. The oldest cold traps should have the highest column abundances (currently estimated as tens of g/cm$^2$).
  
\item The surface might be largely devolatilized and ice exploration methods need to probe the subsurface. As little as 10~cm (approximately the thermal skin depth but also the burial depth inferred from nuclear spectroscopy) should suffice, unless the ice was buried by a landslide or by dry crater ejecta.

\item A key science measurement could be accomplished with a mission to an old (and therefore large) cold trap, where the highest column-abundances of ice are expected. A probe should encounter ice concentrations larger than those seen by LCROSS, and it could also document the contrast between surface and subsurface \h2o concentrations.

\item Massive ice at greater depths?   Existing observations leave room for higher ice concentrations at depths beyond a few meters. From a theoretical perspective, the Cassini State transition and its aftermath (which Mercury did not experience) limit the total amount of water that has been delivered to the cold traps. The standard model does not support deep massive ice deposits.

\item The standard model does {\it not} predict any ice outside of cold traps. Other theoretical explanations are necessary for this. Subsurface cold-trapping, as opposed to surface cold-trapping, could give rise to small concentrations of sequestered \h2o over large areas simultaneously, and should thus be relatively easy to verify or refute. Relic buried ice may be restricted to local burial events. 
  
\item The cold trap areas of supervolatiles HCN, SO$_2$, NH$_3$, CH$_3$OH, and CO$_2$ are much smaller than for \h2o \cite{landis22,sw24}, so most cold-trapped ice deposits should consist purely of \h2o mixed with regolith. The LCROSS impact site was exceptionally cold, which is why supervolatiles were detected there.

\end{itemize}

\vspace{1em}
{\it Acknowledgments.}
This article was made possible by many prolific authors of published research articles. I thank Paul Hayne, David Lawrence, and Matthew Siegler for insightful discussions since the original first version of the manuscript.

\appendix
\section*{Appendix}
\addcontentsline{toc}{section}{Appendix}

\subsection*{A. Acronyms}
\addcontentsline{toc}{subsection}{A. Acronyms}

\begin{tabular}{ll}
  ARTEMIS & two heliophysics spacecraft (2009--now) \\
  %Acceleration, Reconnection, Turbulence, and Electrodynamics of the Moon's Interaction with the Sun \\
  CCGE & Cold Cathode Gauge Experiment on Apollo landers \\
  CHACE & Chandra's Altitudinal Composition Explorer onboard Chandrayaan 1 and 2\\
  ISR  & ice stability region \\
  LCROSS & Lunar Crater Observation and Sensing Satellite \\
  LEND & Lunar Exploration Neutron Detector onboard LRO \\
  LPNS & Lunar Prospector Neutron Spectrometer (1998--1999) \\
  LOLA & Lunar Orbiter Laser Altimeter onboard LRO \\
  LRO & Lunar Reconnaissance Orbiter (2009--now) \\
  NSR & neutron suppression region \\
  M$^3$ & Moon Mineralogy Mapper onboard Chandrayaan 1  (2008--2009) \\
  MESSENGER & Mercury Surface, Space Environment, Geochemistry, and Ranging (2011--2015) \\ 
  Mini-RF & Miniature Radio Frequency onboard LRO \\
  PSR & permanently (or perennially) shadowed region \\
  SELENE & Selenological and Engineering Explorer (2007--2009) \\
  SELENE-TC & SELENE Terrain Camera \\
  WEH & water equivalent hydrogen (for nuclear methods) \\
\end{tabular}

\subsection*{B. Thermodynamic perspective of cold-trapping on airless bodies}
\addcontentsline{toc}{subsection}{B. Thermodynamic perspective of cold-trapping on airless bodies}

The sublimation rate of ice into vacuum is determined by the {\it saturation} vapor pressure of ice by means of the Hertz-Knudsen equation. This may be confusing, because vacuum is not ``saturated''. The conundrum is easily resolved. The saturation vapor pressure of a solid (often also called the equilibrium vapor pressure) is the partial pressure of the gas such that the flux of molecules that impinge and condense on the surface of the solid equals the flux of molecules that leave the surface. At low density, the flux that leaves the surface is the same with or without incoming molecules. The term ``saturation'' is somewhat of a misnomer as this gas pressure really encodes the sublimation energy of the solid.

It is sometimes said that ``\h2o condenses at 110~K''. This was not the original argument given, nor is it true out of context, but with some theoretical abstraction one can make a thermodynamic argument that places the cold trap temperature threshold on the frostline of the phase diagram of \h2o, or at least close to it \cite{schorghofer22stat}. The flux in or out of a cold trap can be assigned a vapor pressure, based on the momentum of the molecules and a time average. Thus even an exosphere has a well-defined pressure, and if one makes the time average long enough, sporadic delivery events are subsumed into that pressure. At a temperature of 109~K, the sublimation rate is 100~\kgmsGa, which corresponds to a pressure of $2\times 10^{-12}$~Pa. In that sense a sublimation threshold corresponds to a frost point temperature. The relation is not perfect, because the temperature inside the cold trap and the temperature of the exosphere (thermalized with the surrounding surface) are different, but it is vaguely true that the cold trap temperature lies on the solid-vapor line of the phase diagram of \h2o, when the vapor pressure is calculated from a very long time average.

\subsection*{C. Fractal photometric paradox}
\addcontentsline{toc}{subsection}{C. Fractal photometric paradox}

The ``fractal photometric paradox'' refers to the idea that the smaller the horizontal resolution, the more shadowed area there is. A small boulder casts a shadow not discernible at coarser resolution. It is no more of a paradox than the idea that an infinite series can converge to a finite value (as in Zeno's paradox). The name derives from the ``photometric paradox'', also known as Olbert's paradox (the darkness of the sky).

Petrov et al. (2003) \cite{petrov03} have applied this concept to PSRs on the Moon, using a hierarchical landscape model with Gaussian surfaces and found the cumulative shadowed area to increase quickly. The cumulative PSR area, measured based on raytracing of LOLA topography, is higher for finer horizontal resolution of the topographic map \cite{smith17,barker23}.
An extensive analysis based of shadows in orbital images and Diviner temperatures showed that micro cold traps extend the global cold trap area by 10--20\% \cite{hayne21}.

The ``paradox'' has a tangible consequence: The existence of many micro cold traps. (Here, ``micro'' refers to any lateral extent smaller than Diviner's resolution, i.e., their temperatures have not been measured from orbit.)
%Along the same line of thinking, one could also introduce ``nano'' cold traps for those that cannot be resolved by orbital cameras.)
The cumulative area becomes only somewhat larger at finer resolutions. But the number of cold traps increases very rapidly. The same can be said for PSRs \cite{barker23}.
Micro cold traps are easy exploration targets. However, they may well be too young to have been around during the most recent water delivery event.

More generally, forming spatial averages over heterogeneous topography leads to a suite of interesting theoretical problems on airless bodies. Due to the lack of atmospheric heat transfer, lateral temperature gradients can be large and ``anisothermality'' is present below the resolution of orbital instruments \cite{sefton-nash19}. Effective descriptions that deal with the unresolved heterogeneity need to be developed. The fractal photometric paradox is one, relatively simple, example of that.
Another is thermal emission from rough surfaces with shadows and obstructed view.

\addcontentsline{toc}{section}{References}
\bibliographystyle{acm}
%\bibliography{/home/norbert/Papers/planet.bib,/home/norbert/Papers/my.bib,/home/norbert/Papers/prop.bib,/home/norbert/Papers/mars.bib}

\end{document}